# Single-shot fluctuations in waveguided high-harmonic generation

**S.J. Goh[1*], Y. Tao[1], P.J.M. van der Slot[1], H.J.M. Bastiaens[1], J. Herek[2],**
**S.G. Biedron[3], M.B. Danailov[4], S.V. Milton[3], K.-J. Boller[1]**

[1]*Laser Physics and Nonlinear Optics,* [2] *Optical Sciences,*
*Mesa+ Institute for Nanotechnology, University of Twente, Enschede, the Netherlands*
[3] *Dept. of Electrical and Computer Engineering, Colorado State University, Fort Collins, Colorado, USA*
[4] *FERMI@Elettra, Sincrotrone Trieste S.C.p.A., Basovizza, Trieste, Italy*
*[*]s.j.goh@utwente.nl*

**Abstract:** For exploring the application potential of coherent soft x-ray (SXR) and extreme ultraviolet radiation (XUV) provided by high-harmonic generation, it is important to characterize the central output parameters. Of specific importance are pulse-to-pulse (shot-to-shot) fluctuations of the high-harmonic output energy, fluctuations of the direction of the emission (pointing instabilities), and fluctuations of the beam divergence and shape that reduce the spatial coherence. We present the first single-shot measurements of waveguided high-harmonic generation in a waveguided (capillary-based) geometry. Using a capillary waveguide filled with Argon gas as the nonlinear medium, we provide the first characterization of shot-to-shot fluctuations of the pulse energy, of the divergence and of the beam pointing. We record the strength of these fluctuations *vs.* two basic input parameters, which are the drive laser pulse energy and the gas pressure in the capillary waveguide. In correlation measurements between single-shot drive laser beam profiles and single-shot high-harmonic beam profiles we prove the absence of drive laser beam-pointing-induced fluctuations in the high-harmonic output. We attribute the main source of high-harmonic fluctuations to ionization-induced nonlinear mode mixing during propagation of the drive laser pulse inside the capillary waveguide.

## 1. Introduction

High-harmonic generation (HHG) is a nonlinear optical process that provides coherent radiation in the form of ultra-short pulses, covering a broad spectrum including the extreme ultraviolet (XUV). The process is typically driven by femtosecond pulses focused to high intensities (in the order of $10^{14}$ W/cm$^2$ [1, 2]) in samples of noble gas, which are usually supplied in the form of gas-jets [2, 3], gas-cells [4-6] or in thin, gas-filled capillaries [7, 8]. Conversion efficiencies reach values of up to $10^{-5}$ to $10^{-6}$, for instance using Argon [9-14].

A fundamental property of HHG associated with the inherent high nonlinearity of the process is that all parameters of the output radiation fluctuate from pulse to pulse, and that the spatial coherence becomes reduced via distortions of the output beam cross section and phase fronts. For applications of HHG it is of central importance to characterize these fluctuations and distortions, such as for increasing the measurement precision for absolute, nonlinear ionization cross sections [15, 16]. Other examples requiring a characterization of fluctuations and distortions include lens-less diffractive imaging with maximum resolution and optimum utilization of the dynamical range [17], or injection seeding at free-electron laser facilities for improving the coherence and the shot-to-shot stability of the laser output [18-20].

In geometries based on free propagation of the drive laser beam, i.e., in gas cells and gas jets, fluctuations in HHG have been characterized extensively, via recording the directional fluctuations [21, 22], fluctuations of the pulse energy [21, 23], and also spectral fluctuations [24]. In contrast to free propagation, one might expect a lower degree of output fluctuations and a higher degree of spatial coherence, when controlling the drive laser propagation through the gas sample via waveguiding. Such control is achieved when providing the gas sample inside a thin capillary, also called hollow core fiber. Here the propagation of the input pulse produced by the drive laser is confined in the form of waveguide modes [25]. Indeed, there are experimental signatures that HHG in capillary waveguides can be spectrally controlled, in the form of dedicated suppression or enhancement of high-harmonic orders [8, 26, 27]. Similarly, expecting reduced beam pointing fluctuations by waveguiding of the input pulse, it has been suggested to use capillary based HHG for injection seeding instead of free propagation of the input pulse [28].

On the other hand, one may also argue that waveguiding enhances nonlinear effects that modify the input pulse propagation to an undesired degree and, correspondingly, also enhances fluctuations in the HH output. For instance, via ionization-induced spectral broadening and plasma-induced refraction the drive pulses can undergo self-compression [14, 29]. Similarly, intensity dependent index changes can distort the wave fronts of the driving laser pulse, thereby exciting propagation in multiple waveguide modes (higher-order modes), the superposition of which creates spatio-temporal hot spots that further distort the drive laser propagation [14, 30]. Eventually, there might also be simple, direct effects. For instance, pointing fluctuations might be turned into intensity fluctuations of the driving laser pulse inside the capillary because a capillary waveguide acts as a spatial mode filter for the incident beam. The signature of such effects should be a correlation of the fluctuations in the HH output with drive laser beam pointing fluctuations.

Surprisingly, to our knowledge, there is no characterization or quantification of shot-to-shot (single-pulse) fluctuations or shot-to-shot output beam deformations in waveguided HHG. Changes of the beam direction and shape have only been addressed by averaging over a large number of pulses [31]. What is rather required is to record the output parameters of single output pulses, for instance the XUV pulse energy, the divergence, or the direction of emission. Then, using series of such single-pulse recordings, one can provide a statistical analysis of pulse-to pulse (shot-to shot) fluctuations, for instance in the form of standard deviations of single parameters, or in the form of a correlation of fluctuations in pairs of output parameters. To enable a comparison with HHG from jets and cells, it appears important to provide a first, basic characterization of the type and size of fluctuations that are present in capillary based HHG.

Here we present the first shot-to-shot characterization of high harmonic generation in a waveguiding geometry. Using Argon, we characterize the directional fluctuations (beam pointing stability) of the HH output, its fluctuations in beam divergence and the fluctuations of the HH pulse energy (energy jitter). To enable a better tracing of possible reasons for fluctuations, we characterize also correlations between various input and output parameters, such as between pointing fluctuations of the drive laser and the HH output.

## 2. Experimental setup

The experimental setup used for HHG is schematically shown in Fig. 1. To generate the drive laser pulses we employ a Ti:Sapphire infrared (IR) laser system (Legend Elite Duo HP USP, Coherent Inc.) with a repetition rate of 1 kHz and a center wavelength of 795 nm. The laser provided almost Fourier-limited pulses with 40 fs duration and a time bandwidth product (TBP) of 0.5 as measured with a home-built Grenouille [32]. The nominal maximum pulse energy of the laser is 8 mJ but, in order to avoid any major self-phase modulation along the path into the gas-filled capillary, we keep the pulse energy below or equal to 1.1 mJ. A rotatable half-wave plate followed by a polarizing beam splitter is used as variable attenuator. For the maximum pulse energy we calculate [33] a maximum increase of the TBP by a factor of 1.3 at the entrance to the gas capillary, in which we included the beam path of nearly three meters through air, the half-wave plate, the focusing lens and the entrance window for transmitting the beam into the vacuum vessel that contains the capillary.

The beam is focused with a lens of 75 cm focal length into a 67 mm long capillary having an inner radius of $a = 75$ μm. This focal length is chosen to match the beam waist radius to the radius of the lowest-order waveguide mode of the capillary [34]. The capillary is mounted in the vacuum vessel at a distance of 50 cm from the entrance window, with the entrance of the capillary in the focal plane of the lens. For filling the capillary with gas in a controlled manner, the capillary is equipped with six 0.4 mm wide slits as shown in the inset of Fig. 1. Two of the slits, located at 5 and 42 mm from the entrance of the capillary, respectively, are used as gas inlets with a continuous flow and define an interaction length, $L_m$ = 37 mm with a constant pressure profile. The remaining four slits are evenly distributed between 45 mm (3 mm downstream the second slit for gas inlet) and 53.4 mm from the entrance of the capillary. These four slits are used for differential pumping. The capillary ends 13.6 mm behind the last pumping slit. The interaction length in this capillary configuration provides wave guiding for the drive laser pulses over about four Rayleigh lengths of the focused beam ($z_R$ = 9 mm). For alignment, each end of the capillary is mounted on a two-axis translation stage. The alignment comprises maximizing the drive laser throughput and simultaneously restricting the drive laser output to lowest-order mode. We measure a maximum capillary throughput of 50 %. This is less than the theoretical limit of 96 % but is comparable with previously reported values [25, 35]. We attribute the deviation from maximum theoretical throughput to several experimental imperfections, including a slightly elliptical beam profile (1:2 aspect ratio), a slightly off-unity beam parameter ($M_x^2$ = 1.3; $M_y^2$ = 1.1), and that part of the guided light is scattered at the various slits in the capillary. Another indicator for the alignment was a visual inspection of fluorescence emitted transversely from the gas in the capillary. This gives a qualitative impression on how homogeneously the drive laser intensity is distributed along the capillary. At lower gas pressures, a homogeneous distribution of fluorescence coincided with a maximum transmission of the drive laser through the capillary.

With the minimum drive laser pulse energy of 0.6 mJ used and an estimate of 30 % losses due to incoupling and scattered light at the first slit, a peak intensity of up to $1.8 \times 10^{14}$ W/cm$^2$ can be launched into the interaction length in the waveguide. Behind the capillary, for reducing the drive laser intensity by diffraction, we let the HH and drive laser beams co-propagate over a distance of 1.5 m. The drive laser beam is then blocked by a set of two 200 nm thick Aluminum (Al) filters placed in series. The filters act as a band pass for XUV radiation, transmitting approximately half of the HH radiation in the wavelength range of 17 nm to 80 nm. The transmitted HH beam is detected with an XUV CCD camera (Andor, DO420-BN) placed 20 cm behind the filters. This camera position is in the far-field of the HH beam, as can be seen from the small Fresnel number, $N_f = a^2/L\lambda$ between 0.04 and 0.2, calculated for the transmitted wavelength range, where $a$ = 75 µm is the radius of the capillary and $L$ = 1.7 m is the distance from the capillary exit to the XUV camera. Due to the low Fresnel number, the divergence of the HH beam and the direction of emission can be straightforwardly obtained from beam profile measurements. Fig. 2(a) shows a typical single-shot measurement of the beam intensity profile. Comparison with a fit curve shows this profile to be near-Gaussian in both transverse directions.

For characterizing the size of shot-to-shot fluctuations in the high-harmonic output, we record series of 100 single shots. The sample size is set on one hand to obtain an error of 10 % or less at 95 % confidence interval for the statistically determined parameters and on the order hand measure subsequent series of shots under equal conditions. These single-shot measurements are performed at a rate of 3 Hz, by blocking about 332 drive laser pulses per third of a second with a triggered combination of an optical chopper and a magnetic shutter. For measurements of the HH spectrum, a high-line-density, home-made transmission grating (10,000 lines/mm) [36] is moved into the HH beam path at 1.7 cm distance in front of the XUV camera. Measurements of average high-harmonic spectra are obtained by letting the camera integrate over 1000 subsequent shots. Fig. 2(b) shows a typical HH spectrum with four harmonic orders ranging from the 17$^{th}$ up to the 23$^{rd}$ order, recorded with an input pulse energy of 0.6 mJ and a gas pressure of 53 mbar. The spectrum is limited on the short-wavelength side to about 34 nm (23$^{rd}$ order). This wavelength agrees well with the calculated cut-off wavelength assuming that 50% of the drive laser energy has reached the end of the capillary which coincides with our measurement of 50% throughput efficiency for the drive laser. On the long-wavelength side the spectrum is limited to about 47 nm (17$^{th}$ order) by strong re-absorption of the generated XUV radiation in Ar [7].

In order to devise an appropriate measurement setup for fluctuations of the HH pulse energy, we considered that such fluctuations may be caused via two different mechanisms. The first is an energy jitter and beam pointing fluctuation of the drive laser [21], which can be called an external effect, letting the amount of pulse energy that is coupled into the waveguide fluctuate. The second mechanism may be called intrinsic, i.e., when fluctuations are rather based on highly nonlinear effects in the capillary such as associated with spatio-temporal reshaping of the drive laser pulse inside the capillary. It has recently been observed that such reshaping is related to beating of drive laser light propagating simultaneously in the fundamental and higher-order waveguide modes, when higher order modes become excited through ionization-induced scattering of the driving laser pulse [14, 30]

To possibly discriminate between the two mechanisms, we capture with a CMOS camera single shots of the drive laser beam profile while, at the same time, the XUV camera captures single shots of the HH beam profile. The CMOS camera is positioned in the residually transmitted drive beam behind the last folding mirror ($M_f$ in Fig. 1) and is triggered for synchronous recording with the XUV camera. Based on single-shot pairs of XUV and CMOS camera images it becomes possible to search for correlations between the drive laser and HH beam energy and pointing fluctuations.

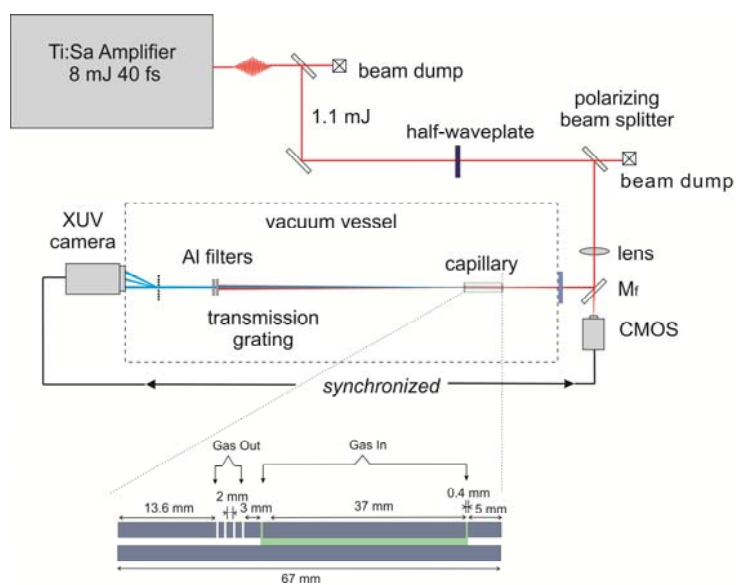

Fig.1. Experimental setup for high harmonic generation (HHG) with a waveguiding, gas-filled capillary.

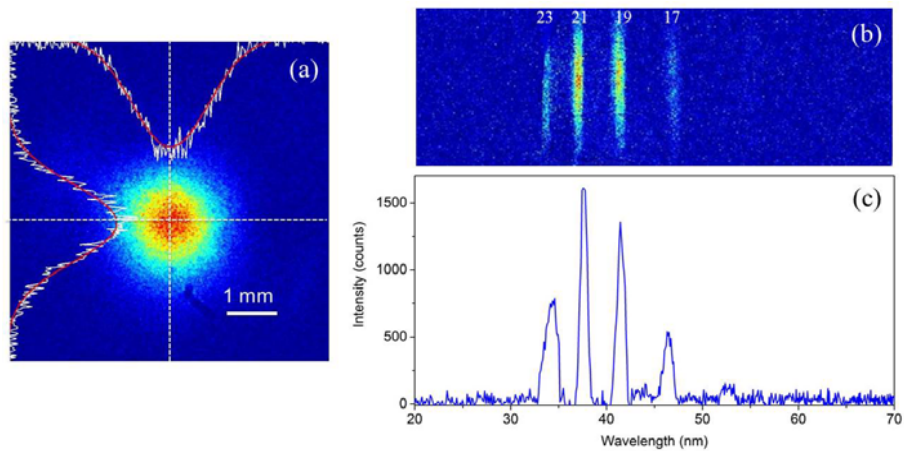

Fig.2. (a) Single-shot XUV camera image of the far-field high harmonic beam cross section showing a near-Gaussian profile: the measured beam profiles (white traces) taken along the respective axes (white dotted lines) matches well with Gaussian fir curves (red). (b) Raw CCD data of high harmonic spectrum (integrated over 1000 shots) with color representing counts, horizontal and vertical axes are pixels. (c) Processed high harmonic spectrum of (b) where intensity is plotted versus calibrated wavelength. Both measurements are taken with a drive laser pulse energy of 0.6 mJ (peak intensity = $1.8 \times 10^{14}$ W/cm$^2$) and an Ar pressure of 53 mbar.

For choosing appropriate settings of the drive pulse energy and the gas pressure in the capillary, we recall that the HH output can be maximized by adjusting the gas pressure to optimize phase-matching [7], while keeping the drive pulse energy constant. The HH output can also be increased, and its spectrum can be extended towards a shorter cut-off wavelength, by increasing the pulse energy of the driving laser pulse (while readjusting the pressure) [37]. Here, we follow a heuristic approach in which we record the high-harmonic output at four different drive pulse energies (between 0.6 and 1.1 mJ) and at six different gas pressures (between 40 and 160 mbar), which are chosen to include the maximum HH output.

### 3. Determining the harmonic pulse energy

To provide an absolute value for the total pulse energy in the high harmonic beam, we combine the data from the measured fluence (beam profile) and spectrum. The background-corrected fluence is summed over the transverse plane to obtain the measured counts, $S$, from the corresponding CCD signal, which is a measure for the total energy in the high harmonic beam. In order to convert $S$ into energy, we need to know the relative contribution of each harmonic, as the conversion from count to energy is wavelength dependent. The spectrum (e.g., see Fig. 2b) is used to determine the relative magnitude, $f_q$, for each harmonic and we assume that the same relative distribution of harmonics is present in the fluence measurement. The contribution, $f_q S$, of each harmonic can now be converted to energy using a wavelength dependent calibration factor, $c(\lambda)$, that takes into account the transmission spectrum of the Al filters and the responsivity of the CCD camera. The energy in a particular harmonic, $E_q$, can be written as

$$E_q = c(\lambda_q) \cdot f_q \cdot S, \tag{1}$$

and the total energy in the high harmonic beam is obtained by summing over all harmonics. The energy calibration factor in our detection (in J/count) is given by $c(\lambda) = E_{eh}\sigma/(\eta_{QE}(\lambda)T(\lambda))$, where $E_{eh}$ is the energy required to generate an electron-hole pair in the silicon detector chip of the CCD camera ($E_{eh}$ =5.84x10$^{-19}$ J $\cong$ 3.65 eV), $\sigma$ is the sensitivity of the CCD camera (10 electrons/count), $\eta_{QE}(\lambda)$ is the specified quantum efficiency of the XUV camera [38], and $T(\lambda)$ is the transmission spectrum of the pair of Al filters. This spectrum is calculated from the CXRO database [39], taking into account the XUV transmission spectrum of aluminum as well as that of thin surface layers of aluminum oxide that are known to form upon contact with oxygen in air. For the calculation of the filter transmission, we assume a layer thickness of 3 nm (on either side) that is typical for filters stored in an oxygen free environment to minimize further oxidation [9, 22]. We note that the aluminum oxide layers reduce the transmission by a factor of 2 to 2.5 across the wavelength range from 30 to 50 nm and therefore have to be taken into account to avoid underestimating the harmonic pulse energy.

Figure 3 summarizes the spectral variation of the quantum efficiency of the XUV camera, $\eta_{QE}(\lambda)$, the filter transmission, $T(\lambda)$, and the calibration factor, $c(\lambda)$. It can be seen that for a wavelength of about $\lambda_{23} = 34$ nm where the strongest harmonic order is found in the measured spectra (the 23$^{rd}$), $c(\lambda_{23}) \approx 1.2 \times 10^{-16}$ J/count. Alternatively expressed, an XUV pulse energy of 1 nJ corresponds to about $8 \times 10^6$ counts for the 23$^{rd}$ harmonic.

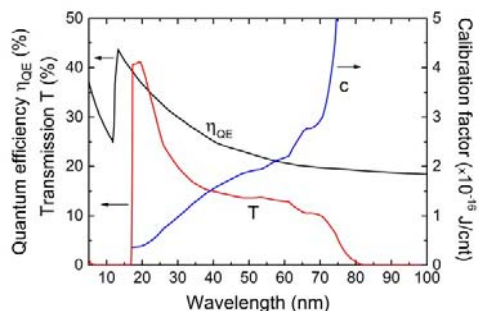

Fig. 3. Typical quantum efficiency ($\eta_{QE}$) of the XUV CCD camera, the transmission ($T$) of the filter set of two Al filters (each 200 nm thick) with an oxide layer on each side (3 nm thick), and the calibration factor ($c$) versus wavelength.

## 4. High-harmonic beam profile, pulse energy and energy jitter

For a first qualitative overview we present in Fig. 4 four typical single-shot CCD images of high-harmonic beam profiles (fluence), recorded with four different drive pulse energies (0.6, 0.8, 1.0, and 1.1 mJ) at the same pressure (53 mbar). This particular pressure was selected because it yielded the highest HH pulse energy. For the lowest drive laser energy of 0.6 mJ we observe that in all shots the HH output exhibits a round, near Gaussian profile such as in Fig. 4a. At higher pulse energies (0.8 mJ and most obvious at 1.0 and 1.1 mJ) the beam profile of each shot shows shot-to-shot fluctuations and becomes increasingly distorted in a complex manner that deviates noticeably from a Gaussian profile. Especially at the two higher pulse energies, the beam profiles vary strongly from shot to shot.

We address this transition from a round and stable profile to a deformed and fluctuating profile to the onset of significant ionization of the Argon gas at drive energies above 0.6 mJ. The spatial variation in ionization (i.e., refractive index) leads to scattering of the input pulse in higher-order modes, while the fast variation of the ionization with time leads to broadening of the spectrum of the input pulse. The spectral broadening and nonlinear mode mixing change the spatiotemporal profile of the driving laser pulse [14]. Note that this is strongly related to ionization-induced defocusing dynamics that can be observed in HHG using free-space propagation of the driving laser pulse [40]. The threshold electron density $n_{th}$ for multi-mode propagation of the driving laser pulse is given by [41]

$$n_{th} = \frac{mc_0^2}{4\pi e^2}\frac{u_{12}^2-u_{11}^2}{a^2} \cong \frac{7}{a^2}\times 10^{12} \text{ cm}^{-3} \qquad (2)$$

where $m$ is the electron mass, $c_0$ is the speed of light in vacuum, $e$ is the electron charge, $u_{11}$ and $u_{22}$ are the first and second zeros of the zeroth-order Bessel function and $a$ is the radius of the capillary (expressed in cm in the numerically evaluated form). The ionization induced electron density can be calculated via the Ammosov-Delone-Krainov (ADK) model [42], showing that the electron density grows with the gas density (gas pressure) and the drive pulse intensity. For the parameters of Fig. 4, eq. 2 predicts a threshold for the electron density of $n_{th} = 1.3\times 10^{17}$ cm$^{-3}$ and from the ADK model we obtain an intra-capillary pulse energy of 0.5 mJ to reach this threshold electron density. Due to in-coupling losses at the entrance of the capillary and scatter losses at the slits, we find that for the lowest input pulse energy (at the entrance of the capillary) of 0.6 mJ the electron density will remain below the threshold, at 0.8 mJ the electron density will be above but near the threshold, and the other two pulse energies are well above the threshold. As stated, we expect that the shape of the IR guided mode field becomes significantly affected in the latter regime [7, 14], and that the phase matching conditions based on the excitation of single fundamental mode propagation [43] will be severely degraded [14]. This should cause the beam profile of the generated HH to deviate noticeably from a round and near-Gaussian shape. At the same time, the strong, optically nonlinear coupling of drive laser light travelling in a number of excited waveguide modes should amplify any small fluctuations of other experimental parameters that would otherwise have only negligible influence. From our experimental observations that HH beam deformation and fluctuations set in where multimode propagation of the driving laser pulse is expected, we conclude that the nonlinear dynamics of the driving laser pulse described in [7, 14, 43] and [29] is also the reason for our observation of shot-to-shot fluctuations. Inducing such dynamics deliberately might increase the HH output significantly [14] but it is less preferred if high-quality XUV beams with low fluctuations are required.

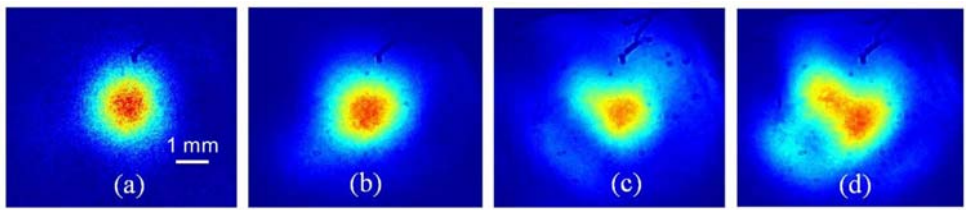

Fig. 4. Typical single-shot images of the high-harmonic beam profile (fluence) *vs.* increasing drive laser pulse energy at a gas pressure of 53 mbar: (a) 0.6 mJ (b) 0.8 mJ (c) 1.0 mJ (d) 1.1 mJ, measured with an XUV CCD at a distance of 1.7 m from waveguide capillary.

To identify the experimental parameters that yield a maximum HH output we record the output for a number of different gas pressures and drive laser pulse energies. Figure 5a shows the total (spectrally integrated) HH output pulse energy averaged over 100 single shots versus pressure for different drive laser energies. It can be seen that for all four values of the drive pulse energy, the maximum average pulse energy is obtained at around 53 mbar of pressure, which we address to optimum phase matching at this pressure. For the 0.6 mJ input pulse energy, where single-mode propagation of the driving laser pulse is expected, we find that the observed pressure of 53 mbar for optimum phase matching agrees well with the value predicted by the single-mode model of Constant *et al.* [44] for our parameters.

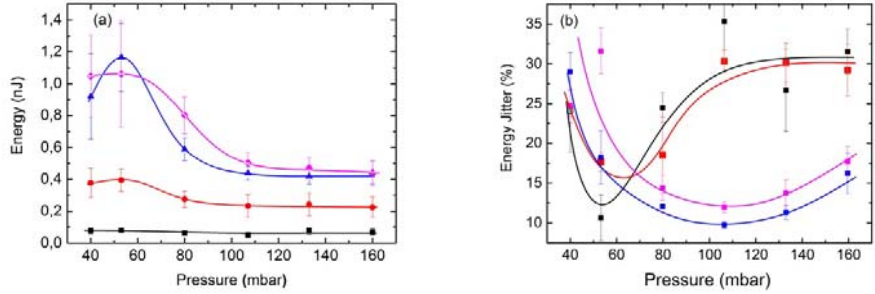

Fig.5. (a) Total (spectrally integrated) pulse energy of the high-harmonic beam measured at the XUV camera as a function of Argon pressure for different drive laser energies at 0.6 mJ (black square), 0.8 mJ (red circles), 1.0 mJ (blue triangles), and 1.1 mJ (purple diamonds). Each data point is obtained via measuring the single-shot pulse energy for 100 shots and subsequently calculating the average value and standard deviation (error bars). (b) Relative energy jitter, defined as the standard deviation normalized to the average pulse energy.

For quantifying the fluctuations of the output energy (energy jitter), we plot in Fig. 5b the standard deviation of the measured pulse energy normalized to its average. This provides the relative energy jitter of high-harmonic generation as a function of gas pressure. It can be seen that, for the two lower drive laser pulse energies (0.6 and 0.8 mJ), the jitter is lowest at 53 mbar (12% and 20%, respectively), where also the average output is highest. As discussed before, this is where phase matching with the fundamental waveguide mode occurs. A possible explanation for lowest fluctuations at best phase matching is as follows. At phase matching, to first order and within the phase-matching bandwidth, the HH output does not vary with any parameter that enters into the wave-vector balance. For instance, if there are shot-to-shot fluctuations in the degree of ionization, phase matching would remain fulfilled to first order and the HH output would not fluctuate as strongly. When the pressure is tuned away from the optimal value for phase matching we observe a simultaneous decrease in HH yield and increase in relative fluctuations. Due to the higher pressure, the electron density surpassed the threshold, even at the lowest input energy. We therefore expect multimode propagation for the driving laser pulse at these pressures and therefore a higher relative fluctuation as the HHG process becomes more susceptible to small perturbations.

The two higher input pulse energies (1.0 and 1.1 mJ) show a different behavior. For these input pulse energies high fluctuations are found near the maximum HHG output at 53 mbar, while the fluctuations are strongly reduced for higher pressures. High fluctuations are expected for multi-modal propagation of the driving laser pulse, and this observation is consistent with the observed distortions in the beam profiles at a pressure of 53 mbar (see Figs. 4b and c). For both pulse energies, the lowest fluctuations occur around a pressure of 106 mbar (see Fig. 5b) where the HH yield is lower. It can also be seen that the range of low fluctuation is rather wide, reaching almost 100-mbar width for 1-mJ input pulse. At the same time we observe that in this range the beam profile is very similar to that shown in Figs. 4a and b and stable from shot-to-shot. This suggests that under these conditions, i.e., high pressure and high pulse energy, the complex, ionization-induced nonlinear dynamics creates a single propagating mode for the driving laser pulse [30]. Although this mode will be very different from the fundamental (empty) waveguide mode, it will produce stable and reproducible XUV output in a similar way.

Our results shows that, depending on the pressure and drive laser intensity, there is a best compromise between achieving the highest pulse energy and the lowest energy jitter. This should be taken into consideration when designing a HH source for specific application.

## 5. Measurement of harmonic beam divergence and pointing stability

To characterize the spatial beam fluctuations, we first measure the average divergence of the HH output versus pressure and drive laser pulse energy. We define the half-angle divergence as the ratio of the beam radius to the axial distance ($L$) of the XUV CCD camera ($L$=1.7 m). The beam radius is taken as $2\sigma_x$ ($2\sigma_y$), where $\sigma_x$ ($\sigma_y$) is the standard deviation of the background-corrected fluence distribution in the $x(y)$-direction. The divergence is calculated for 100 shots separately and then both the average and standard deviation are calculated for each set of 100 shots. In Fig. 6 we plot the average and the standard deviation of the divergence in both the horizontal (Fig. 6(a)) and vertical (Fig. 6(b)) direction versus pressure for different input pulse energies. The error bars represent the standard deviation. It can be seen that the horizontal beam divergence lies in the range of 1.2 to 1.6 mrad and shows only a weak increase with pressure and with pulse energy. The vertical beam divergence shows a stronger variation with pressure and is found to be in the range of 1.3 to 2.1 mrad (Fig. 6(b)). Finally, it can also be seen that the horizontal beam divergence is generally less than the vertical divergence for the collection of input pulse energies and Ar gas pressures investigated.

A dependency on the energy of the driving laser pulse is to be expected as, in first-order approximation, the induced single-atom dipole phase is proportional to the instantaneous intensity of the driving laser pulse. The intensity of the driving pulse inside the capillary is itself a result of a highly nonlinear dynamical process [7,12,14,29,42], as described above. It is therefore surprising that the observed divergence only shows a weak dependence on input pulse energy and gas pressure. However, using a modified multimode generalized nonlinear Schrödinger equation, Anderson *et al.* numerically showed that the size of the most intense part of the driving laser pulse and its intensity only weakly depend on beam energy and gas pressure for most of the parameter range we investigated [30], which is consistent with the weak dependence of the measured divergence on input pulse energy and gas pressure. The half-width at half of the maximum intensity of the driving laser pulse is predicted by their model to be around 10 µm, which agrees well with the apparent source radius $r_a \approx 8$ µm calculated from $r_a = \frac{\lambda}{\pi\theta}$, where $\lambda = 38$ nm is the wavelength of the strongest harmonic and $\theta \approx 1.5$ mrad is an average value for the measured far field divergence (cf., Fig. 6). The difference in horizontal and vertical divergence may be due to the slight elliptical shape of the input beam, leading to a slight asymmetry in the nonlinear dynamics for the driving laser pulse inside the capillary and therefore a slightly elliptical apparent source for the XUV radiation.

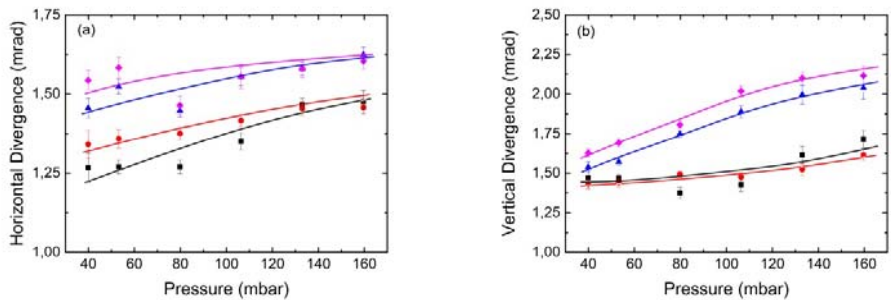

Fig.6. Average beam divergence of the high harmonic output measured as a function of gas pressure for different drive laser intensities at 0.6 mJ (black squares), 0.8 mJ (red circle), 1.0 mJ (blue triangles) and 1.1 mJ (purple

diamonds): (a) for the horizontal direction and (b) for the vertical direction. The error bars represent the standard deviation.

For quantifying the first-order spatial fluctuation of the HH beam, we define the relative beam pointing fluctuation in the horizontal (vertical) direction, $b_x$ ($b_y$), as the standard deviation of the center-of-gravity, $s_x$, of the fluence distribution in the $x(y)$-direction, as measured from 100 shots, normalized to the average horizontal(vertical) beam radius, $\overline{2\sigma_{x(y)}}$,

$$b_{x(y)} = \frac{s_{x(y)}}{2\sigma_{x(y)}}. \tag{3}$$

In Fig.7, we plot the measured horizontal (Fig.7a) and vertical (Fig.7b) pointing fluctuation (pointing instability) versus the pressure for different drive laser energies. It can be seen that the horizontal beam pointing fluctuation lies in the range of 5 to 10 % and does not change significantly with pressure. The vertical beam pointing fluctuation is larger, on the order of 10 to 30%. Yet, with increasing pressure, the beam pointing fluctuation decreases. If we combine these results with the divergence measurements, we find that the absolute pointing fluctuation (i.e., $s_{x(y)}$) is approximately constant and independent of pressure and energy. From this we conclude that the interplay of nonlinear dynamics and capillary waveguiding, providing a stable mode of propagation for the drive laser as described above, indeed provides mode stability over wider pressure and energy ranges as predicted [30].

The observed absolute value for vertical pointing fluctuations (0.3 mrad) is about three times larger than the value for the horizontal fluctuation (0.1 mrad). We attribute this again to the slightly elliptical shape of the input driving pulse.

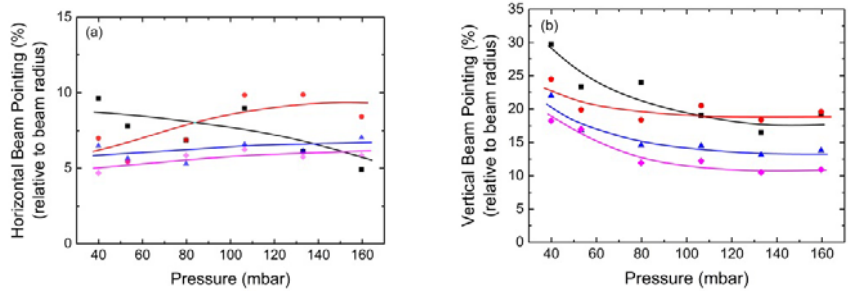

Fig.7. The high harmonic beam pointing fluctuation relative to the average beam radius as a function of gas pressure for different drive laser intensities at 0.6 mJ (black squares), 0.8 mJ (red circles), 1.0 mJ (blue triangles), and 1.1 mJ (purple diamonds): (a) for the horizontal direction and (b) for the vertical direction

In summary, high-harmonic generation in our gas-filled capillary achieves a highest HH output energy of 1.2 nJ at a drive energy of 1 mJ and a gas pressure of 53 mbar. At these conditions, we measure the following beam parameters: an energy jitter of 18 %, a divergence of 1.5 mrad ($x$) and 1.6 mrad ($y$), and a beam pointing fluctuation of 6 % ($x$) and 17 % ($y$). In comparison, non-waveguided HHG as reported by others [21] showed an energy jitter of 6 %, a divergence of 0.7 mrad ($x$) and 0.2 mrad ($y$), and a beam pointing fluctuation of 4 % ($x$ and $y$). In that case the origin of beam pointing and energy jitter was reported as possibly related to drive laser instabilities. Our result shows higher HH output fluctuations although our drive laser fluctuations are comparable with what was reported for the non-waveguided case. This raises the question whether the capillary amplifies the drive laser pulse fluctuations and correlations are present between the fluctuations in the driving laser pulse at the input of the capillary and the fluctuations in the high harmonic output beam.

## 6. Correlation measurement of beam parameters between driver laser and HH

To investigate the influence of drive laser beam fluctuations on HH beam fluctuations, we make use of the single-shot drive laser beam profiles that were simultaneously recorded with the single-shot high-harmonic beam profiles. To reveal a possible correlation between the two, we compared the fluctuations in the center-of-gravity for the two beams (Fig. 8), the HH output energy versus the center-of-gravity of the drive laser beam (Fig. 9) and the energy in the pulses in the two beams (Fig. 10). To quantify the strength of a possible correlation, we calculate the correlation coefficient, $\rho$, between the two data sets, where $\rho = 1$ (-1) indicates a maximum correlation (anti-correlation) and $\rho = 0$ indicates the absence of any correlation, i.e., mutually independent fluctuations.

A typical example of the center of gravity of the HH beam vs. the center of gravity of the drive laser beam is shown in Fig. 8, for each of the 100 single shots at a drive laser energy of 1.0 mJ and at 53 mbar Ar gas pressure, which corresponds to the highest HH output measured. The distributions shown in Fig. 8 are, on a first glance, uncorrelated. Indeed, the correlation coefficients are found to be close to zero; 0.07 and -0.11 for horizontal and vertical fluctuations, respectively. The correlation plots between mutually orthogonal directions [horizontal (vertical) drive laser fluctuations vs. vertical (horizontal) HH fluctuations] are not shown here, because the results are similar, i.e., a correlation of 0.09 for horizontal drive laser beam and vertical HH fluctuations, and a correlation of 0.08 for vertical drive laser beam and horizontal HH fluctuations. The weak beam pointing correlation between the drive laser and HH output beams indicates that drive laser beam pointing fluctuation is not a significant factor for HH beam pointing fluctuation in a waveguided geometry.

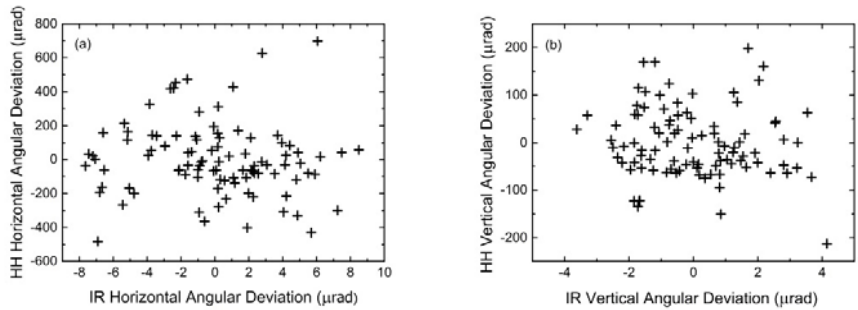

Fig.8. Correlation plot of 100 single-shot measurements of the deviation of the HH beam center of gravity from its average versus the drive laser center of gravity from its average for (a) both horizontal and (b) both vertical fluctuations; the drive laser energy is 1.0 mJ and the Ar gas pressure is 53 mbar. The correlation coefficients are -0.07(a) and -0.11(b), respectively, too small for indicating any correlation with statistical significance.

In Fig. 9 we show the measured HH output energy as a function of the IR beam pointing for 100 single shots. The correlation coefficients are found to be 0.03 and 0.05 for horizontal and vertical pointing fluctuations, respectively. This means again that there is only a weak correlation between the HH output energy and the drive laser beam pointing.

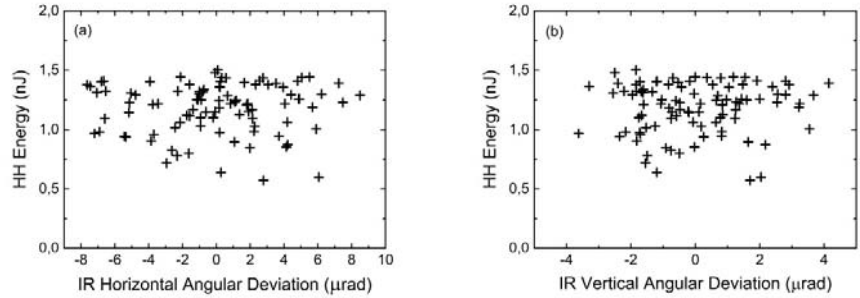

Fig.9. Energy in the HH beam versus the horizontal (a) and vertical (b) deviation of the IR beam center of gravity for 100 single shots. The drive laser energy is 1.0 mJ and the Ar gas pressure is 53 mbar. The correlation coefficients are 0.03 (a) and 0.05 (b), respectively.

Besides directional fluctuations, there is another type of drive laser instability that may cause HH energy fluctuations, which is fluctuation of the drive pulse energy. The presence of such effect should be visible as a correlation between drive laser pulse energy and HH output energy fluctuations. For an analysis, we have plotted the pulse energy of 100 single HH output pulses vs. the energy of the corresponding drive pulses, as seen in the example of Fig. 10. The correlation coefficient obtained from the data is close to zero again ($\rho = 0.11$). We conclude that there are only weak correlations between drive energy and HH energy fluctuations.

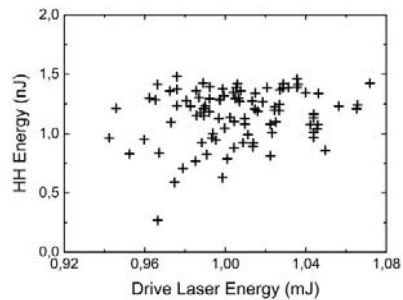

Fig.10. The HH beam energy versus the drive laser energy for 100 single shot. The drive laser energy is 1.0 mJ and the Ar gas pressure is 53 mbar. The correlation coefficient obtained from the data is 0.11.

As a final check for externally induced fluctuations of the HH output we investigated whether there are any instabilities introduced by the capillary. Although the capillary is firmly mounted at an optical table, it might be possible that small turbulences in the gas flow cause small fluctuations in the refractive index of the gas. Similarly, small acoustic perturbations may be guided from the vacuum pump system to the capillary. If significant, these effects should become visible as increased pointing and energy fluctuations of the drive laser behind the capillary. The data on the drive laser presented above shows that the beam pointing and pulse energy fluctuation before entering the capillary are in the range of 0.3-0.5 % and 0.5%, respectively. We have measured the drive laser pointing and energy fluctuations behind the capillary, with the gas flow on and off, and with the vacuum pumps on and off. We observe that there is no difference in fluctuations, with the gas flow and vacuum pumps on or off. However, we still measure that the pointing and pulse energy fluctuations increase to about 3 % for the transmitted drive laser pulse behind the capillary.

The correlation measurements show that there is only a weak correlation between the fluctuations in the drive laser beam and high harmonic beam. Still, we observe higher fluctuations in the transmitted drive laser beam, which do not seem to depend on the gas flow nor on acoustic perturbations from the vacuum pumps. We therefore conclude that the increase in fluctuations in the transmitted beam are due to fluctuations in the coupling of the drive laser beam into the capillary caused by a combination of a slight elliptical shape, a slight off-unity beam parameter and the energy and pointing fluctuations of the driving laser pulse. That we observe only a weak correlation between the drive laser beam and the HH output beam fluctuations means that the propagation through the capillary introduces a randomization that removes any possible correlation. For example, local fluctuations in gas density result in local electron density variation and this effects both the modal spatial and phase distribution. However, a detailed investigation into this is beyond the scope of the present investigation.

## 7. Summary and conclusion

In this work we present what is to our knowledge the first single-shot analysis of the output beam properties and stability in waveguided high-harmonic generation (HHG). The experiments are carried out in a thin waveguiding capillary of standard radius (75 µm inner radius) filled with Ar gas, which yields a maximum HH pulse energy of 1.2 nJ in the range between 29 nm and 52 nm (15$^{th}$ and 27$^{th}$ harmonic of a Ti:Sapphire laser) at a drive pulse energy of 1 mJ at 40 fs drive pulse duration. We experimentally characterize the strength of various types of fluctuations in the high harmonic output, i.e., fluctuations in beam shape, beam pointing, and HH pulse energy vs. the gas pressure and drive laser pulse energy. We attribute the observed effects to ionization- induced nonlinear mode mixing and spectral broadening of the drive laser inside the capillary waveguide. In single-shot correlation measurements we observe a weak correlation between drive laser pointing or energy fluctuations with pointing or energy fluctuations in the HH output. Our investigations show that in waveguided HHG it is essential to have a very stable drive laser beam with very good beam quality and at the same time apply appropriate pressure and drive laser pulse energy for achieving the best compromise between a high beam stability and a maximum output pulse energy.

**Acknowledgments**

This research is supported by the Dutch Technology Foundation STW, which is part of the Netherlands Organization for Scientific Research (NWO), and which is partly funded by the Ministry of Economic Affairs. Additionally support comes from FERMI@Elettra and Coherent Europe BV.